# Modulation of sensory information processing by a neuroglobin in *C. elegans*

Shigekazu Oda[1a], Yu Toyoshima[b] and Mario de Bono[1a]

[a]MRC Laboratory of Molecular Biology, Francis Crick Avenue, Cambridge CB2 0QH, UK;

[b]Department of Biochemistry and Biophysics, University of Tokyo, 2-11-16 Yayoi, Bunkyo-ku, Tokyo, Japan.

[1]To whom correspondence should be addressed. E-mail: debono@mrc-lmb.cam.ac.uk or

soda@mrc-lmb.cam.ac.uk

**Abstract**

Sensory receptor neurons match their dynamic range to ecologically relevant stimulus intensities. How this tuning is achieved is poorly understood in most receptors. We show that in the *C. elegans* URX $O_2$ sensing neurons two putative molecular $O_2$ sensors, a neuroglobin and $O_2$-binding soluble guanylate cyclases, work antagonistically to sculpt a slowly rising sigmoidal $O_2$ response curve tuned to approach saturation when $O_2$ reaches 21%. *glb-5* imposes this sigmoidal function by inhibiting $O_2$-evoked $Ca^{2+}$ responses in URX when $O_2$ levels fall. Without GLB-5, the URX response curve approaches saturation at 15% $O_2$. Behaviorally, GLB-5 signaling broadens the $O_2$ preference of *C. elegans* while maintaining strong avoidance of 21% $O_2$. Our computational aerotaxis model suggests that the relationship between GLB-5-modulated URX responses and reversal behavior is sufficient to broaden $O_2$-preference. Thus, a neuroglobin can shift neural information coding leading to a change in perception and altered behavior.

**Introduction**

Neuroglobins are members of the globin family of oxygen ($O_2$)-binding heme proteins expressed mainly in neurons (Burmester and Hankeln, 2014). They have been described throughout metazoa, from cnidarians to man (Burmester et al., 2000). Their physiological functions are poorly understood but they are proposed to protect neurons from hypoxia by metabolising reactive oxygen species (ROS), redox sensing or signalling, $O_2$ storage, control of apoptosis, and as negative regulators of Gi/o signalling (Burmester and Hankeln, 2009). In the nematode *Caenorhabditis elegans* the neuroglobin GLB-5 (*GLOBIN-5*) enables animals returned to normoxia after prolonged hypoxia exposure to accumulate rapidly where their bacterial food is thickest (Gross et al., 2014). GLB-5, like vertebrate neuroglobins, has a hexa-coordinated heme iron, and rapidly oxidizes to the ferric state in normoxia (Persson et al., 2009). GLB-5 also plays a role in animals not exposed to hypoxia, modifying the response of specialized $O_2$-sensing neurons (Persson et al., 2009; McGrath et al., 2009). While phenomenological change in neural responses by GLB-5 has been reported, how GLB-5 alters the representation of environmental information (neural coding) in these neurons leading to behavioural change is unknown.

*C. elegans* avoids both high and low $O_2$ environments (Gray et al., 2004). Avoiding high $O_2$ helps the animal escape surface exposure, and is mediated by sensory receptors that exhibit phasic-tonic responses, most importantly the URX neurons (Busch et al., 2012; Zimmer et al., 2009). URX, which express the GLB-5 neuroglobin (Persson et al., 2009), evoke both transient behavioral responses that are coupled to the rate of change of $O_2$, $dO_2/dt$, and more persistent responses coupled to $O_2$ levels, [$O_2$]. The transient responses are reversals and turns that allow *C. elegans* to navigate $O_2$ gradients. The sustained responses involve a persistent switch to rapid movement as feeding animals seek to escape 21% $O_2$. Besides *glb-5*, putative molecular $O_2$ sensors in URX include orthologs of the mammalian NO-binding soluble guanylate cyclases called GCY-35 and GCY-36 (Gray et al., 2004; Cheung et al., 2004; Couto et al., 2013). The heme – nitric oxide/oxygen (H-NOX) binding domain of GCY-35/GCY-36 is thought to stimulate cGMP production upon binding molecular $O_2$. The domesticated reference strain of *C. elegans,* N2, harbors a defective *glb-5* allele, but natural isolates encode a functional *glb-5(Haw)* allele (Persson et al., 2009). Here, we show that GLB-5(Haw) and the soluble guanylate cyclases work antagonistically to confer on URX a slowly rising sigmoidal $O_2$-stimulus–response curve tuned to saturate as $O_2$ levels approach 21%. The GLB-5 retuning of URX broadens the range of $O_2$ environments preferred by *C. elegans*. Using computer modelling we show that this altered preference can be explained by changes in how URX evokes reversals in response to $O_2$ stimuli.

**Results**

**The GLB-5 allele in wild-caught strains broadens *C. elegans'* $O_2$ preference**

To analyse how the functional *glb-5* allele found in natural *C. elegans* strains alters $O_2$ preference we compared the distribution of *glb-5(Haw)* and *glb-5(tm5440)* animals in a 0% – 21% $O_2$ gradient in the presence of food. The *glb-5(tm5440)* mutation deletes much of *glb-5* and is predicted to be a null allele. In all our experiments we used strains defective in the neuropeptide receptor *npr-1*, since besides harbouring a defective *glb-5* allele, the *C. elegans* lab strain, N2 (Bristol), has $O_2$-sensing defects due to a gain-of-function mutation in this receptor (Gray et al., 2004; de Bono and Bargmann, 1998). *glb-5(tm5440)* animals accumulated in a narrow range of $O_2$ concentrations, between 7% and 10 % $O_2$ (Fig. 1*A* and *B*). By contrast, animals bearing the natural *glb-5(Haw)* allele distributed over a broader range of $O_2$ concentrations, between 17% and 5% $O_2$, but still avoided 21% $O_2$ and hypoxia (Fig. 1*A* and *B*). These behavioral data imply that *glb-5(Haw)* changes how $O_2$-sensing neurons decode $O_2$ gradients.

**GLB-5 changes the dynamic range of the URX $O_2$ sensor**

We used the GCaMP6s $Ca^{2+}$ sensor to examine how *glb-5* function alters neural coding in the URX $O_2$-sensing neurons (Chen et al., 2013). The dynamics of the $Ca^{2+}$ responses evoked in URX by a 7% – 21% $O_2$ single step stimulus did not differ significantly between *glb-5(tm5440)* and *glb-5(Haw)* animals (Fig. S1). However, the $Ca^{2+}$ responses to a 7% – 19% $O_2$ exponential ramp stimulus differed markedly between these strains (Fig. 1*C*). In animals expressing *glb-5(Haw)*, $Ca^{2+}$ in URX increased continuously as $O_2$ levels rose from 7% to 21%. By contrast, the $Ca^{2+}$ responses of *glb-5(tm5440)* mutants appeared to saturate at ~14 % $O_2$. These differences suggest that the GLB-5 neuroglobin changes the dynamic range of URX, an observation we also made in PQR, another $O_2$ sensor that expresses GLB-5 (Fig. 1*C*). The effect of *glb-5* alleles on the URX $Ca^{2+}$ response was similar whether we imaged animals in the presence (Fig. 1*C*) or absence of food (Fig. S2). As expected, animals defective in both *glb-5* and the *gcy-35* soluble guanylate cyclase URX neurons did not respond to the $O_2$ stimulus (Fig. 1*C*).

To investigate further how the GLB-5 neuroglobin alters neural coding, we delivered different patterns of $O_2$ stimuli and imaged $Ca^{2+}$ responses in URX. We focused on URX because these sensory neurons are sufficient for several $O_2$-coupled behaviors including aerotaxis (Fig. S3) and aggregation (Cheung et al., 2005; Rogers et al., 2006). To plot the relationship between stimulus intensity and URX $Ca^{2+}$ responses we sequentially increased the $O_2$ stimulus given to the same animal in 2% increments, returning to 7% $O_2$ between stimuli (Fig. 2*A*). In *glb-5(Haw)* animals URX showed a higher $O_2$-response threshold than *glb-5(tm5440)* mutants, as well as a steeply sigmoidal $O_2$ response curve whose half maximum was at much higher $O_2$ concentrations (Fig. 2*A* and *B*). Selectively expressing a *glb-5(Haw)* transgene in URX in *glb-5(tm5440)* mutants conferred a URX stimulus–response profile that closely resembled that of *glb-5(Haw)* animals (Fig. 2*A* and *B*). Thus, GLB-5(Haw) cell-autonomously shifts the URX stimulus-response curve towards higher $O_2$ concentrations (Fig. 2*B*).

To extend these observations, we examined URX responses to a different set of stimuli in which we increased $O_2$ levels by 2% but varied the starting $O_2$ concentration and delivered only one stimulus per animal. Again, we observed that the *glb-5(Haw)* allele shifted the dynamic range of URX such that both tonic $Ca^{2+}$ levels, and the change in $Ca^{2+}$ normalized to the pre-stimulus $Ca^{2+}$ level, $\Delta R/R_o$, a measure of response amplitude, gradually increased as $O_2$ approached 19% - 21% (Fig. 3*A* and *B*). By contrast, in *glb-5(tm5440)* mutants $\Delta R/R_o$ was at a maximum when animals experienced an 11→13 % $O_2$ stimulus (Fig. 3*A* and *B*). Expressing *glb-5(Haw)* cDNA selectively in URX in *glb-5(tm5440)* animals was sufficient to confer a *glb-5(Haw)*-like dose-response curve to this neuron (Fig. 3*A* and *B*).

Step stimulation is used widely to study the properties of sensory neurons, but in their natural environment *C. elegans* likely also encounter slowly varying $O_2$ levels, similar to those encountered by animals in the aerotaxis assay (Fig 1*A*). We therefore measured URX $Ca^{2+}$ responses evoked by a set of 2% $O_2$ exponential ramp stimuli. Our results showed a similar pattern to the one we observed for the corresponding step stimulus paradigm (Fig. 3*C* and *D*). In animals expressing the *glb-5(Haw)* allele, URX responses to ramp stimuli increased gradually as $O_2$ levels increased. By contrast, in *glb-5(tm5440)* mutants the URX response amplitudes, measured as $\Delta R/R_0$, showed a peak response to the 13→15 % $O_2$ stimulus and were otherwise similar across the different ramp stimuli we delivered (Fig. 3*C* and *D*). The response property changes conferred by GLB-5(Haw) are therefore robust to different $O_2$ stimulation patterns. Together, our $Ca^{2+}$ imaging experiments suggest that GLB-5(Haw) alters neural encoding of $O_2$ levels in URX, increasing the dynamic range of URX and shifting it to higher $O_2$ concentrations.

**cGMP signalling in URX**

In previous work we used the genetically-encoded cGMP sensor cGi500 (Russwurm et al., 2007) to visualize cGMP dynamics in the PQR $O_2$ sensing neuron (Couto et al., 2013). We showed that a rise in $O_2$ stimulates a tonic GCY-35-dependent rise in cGMP, and that the $Ca^{2+}$ influx resulting from gating of cGMP channels feeds back to limit $O_2$-evoked rises in cGMP by stimulating cGMP hydrolysis (Couto et al., 2013) (Fig. S4*A*). We used cGi500 to examine if GLB-5(Haw) can modulate cGMP dynamics in URX. We could not detect $O_2$-evoked cGMP responses in the cell body of URX neurons unless we disrupted *cng-1*, which encodes a cGMP-gated channel subunit required for $O_2$-evoked $Ca^{2+}$ responses in URX (Fig. S4*A* and *B*). This suggests that URX and PQR have similar negative feedback control of cGMP accumulation. The cGMP responses evoked in URX by an exponential ramp $O_2$ stimulus was comparable in *glb-5(tm5440); cng-1* and *glb-5(Haw); cng-1* animals in our experimental conditions (Fig. 4*B* and *C*). These results suggest that GLB-5(Haw) does not modulate URX neural coding by modulating cGMP levels. However, we cannot exclude the possibility that measuring cGMP in the cell body does not adequately report cGMP changes in the dendritic ending, where GLB-5, GCY-35/GCY-36 and the cGMP channels are localized.

**Ectopic GLB-5 expression can alter the sensory properties of a $CO_2$ sensor**

To explore the ability of GLB-5(Haw) neuroglobin to modify the properties of other neurons, we expressed it ectopically in the AFD sensory neurons. AFD neurons respond to both temperature (Kimura et al., 2004) and $CO_2$ concentrations changes (Bretscher et al., 2011; Kodama-Namba et al., 2013) via mechanisms that involve cGMP signalling. Animals expressing a *pAFD::glb-5(Haw)* transgene showed a larger transient decrease in $Ca^{2+}$ upon exposure to increasing $CO_2$ concentrations than non-transgenic controls (Fig. S5). These results suggest that *glb-5(Haw)* can actively alter neural encoding independently of the $O_2$ sensing mechanisms in URX. We next asked if GLB-5(Haw) expression can make AFD responsive to changes in $O_2$, by imaging AFD $Ca^{2+}$ while delivering $O_2$ stimuli. We found that GLB-5(Haw) conferred $O_2$ responsiveness to AFD, although responses were small (Fig. S6). These results suggest that *glb-5(Haw)* neuroglobin can actively alter neural properties.

**GLB-5 effects on URX behavioral outputs**

How do the changes in URX information coding mediated by GLB-5(Haw) alter motor responses to $O_2$ stimuli? To address this question, we quantified behavioral responses to a range of O$_2$ stimuli, focusing on reversal and speed, which important features of O$_2$-evoked behaviors (Cheung et al., 2005; Rogers et al., 2006). Avoidance of high O$_2$ is mediated principally by three sensory neurons, URX, PQR and AQR, each of which express the GCY-35/36 O$_2$ receptor and GLB-5 (Cheung et al., 2004; Cheung et al., 2005; Gray et al., 2004; Busch et al., 2012). To focus on the behavioral consequences of URX output we studied *gcy-35(ok769); glb-5(tm5440)* mutants that expressed *glb-5(Haw)* and / or *gcy-35* cDNA in URX but not AQR or PQR (see Methods). In these animals the relationship of URX output to reversals showed an unexpected Goldilocks effect (Fig. 4*A* and *B*). O$_2$ stimuli that evoked intermediate Ca$^{2+}$ responses in URX in our imaging experiments evoked strong reversals (Fig. 3*A*, *B*, 4*A* and *B*). However, O$_2$ stimuli that evoked small or large Ca$^{2+}$ responses in URX failed to evoke reversals (Fig. 3*A*, *B*, 4*A* and *B*). The effect of the *glb-5* genotype on reversals was consistent with its effect on the magnitude of the O$_2$-evoked URX Ca$^{2+}$ response (Fig. 3*A*, *B*, 4*A* and *B*). These data suggest that URX associated circuits include a filter that prevents strong stimulation of URX from inducing reversals. Consistent with this, a 13 → 21 % O$_2$ step stimulus that evoked a large URX Ca$^{2+}$ response did not evoke reversals (Fig. S7). URX promoted faster movement only when O$_2$ levels rose above 17% O$_2$ (Fig. 4*C* and *D*). Together, our results suggest that information from URX is transmitted to both reversal and speed circuits, however, the transmission efficiency from URX to the reversal circuit is higher.

**A computational model for aerotaxis**

Using our detailed analyses of how URX responds to different O$_2$ stimuli we performed computational experiments to ask if the relationship between URX activity and reversal behavior could explain the altered aerotaxis preference of animals expressing *glb-5(Haw)*. To build a computational model for aerotaxis we used the $Ca^{2+}$ imaging data in Fig. 3*A* and the behavioral data in Fig. 4*C* (see methods). In the model the position of a worm was represented as a single point, a command neuron randomly generates reversal, and a signal is transmitted from a URX model neuron to the command interneuron via an interneuron (differentiator) to promote reversal (Fig. 5*A*, see Methods). In the URX model, URX $Ca^{2+}$ responses to $O_2$ stimuli were approximated by a Nonlinear-Linear-Nonlinear (NLN) model (Fig. 5*A*). The parameters for the NLN model were estimated from imaging URX responses to 2% step $O_2$ stimuli in *glb-5(tm5440)* and *glb-5(Haw)* (Fig. S8). A single model reproduced URX $Ca^{2+}$ responses to a variety of $O_2$ changes. As a result of modeling the URX responses, we acquired two sets of parameters, one for *glb-5 (tm5440)* and the other one for *glb-5 (Haw)*. The parameters for steps downstream of URX were common for *glb-5 (tm5440)* and *glb-5 (Haw)* virtual animals.

    Having set up our model we ran *in silico* aerotaxis experiments. These experiments showed that worms in which the URX NLN model used parameters obtained for *glb-5(m5440)* preferred 7% – 10 % $O_2$ (Fig. 5*B*, *C* and S9) whereas those using *glb-5(Haw)* parameters showed broader $O_2$ preference, with the majority of worms preferring 7% – 16 % $O_2$ (Fig. 5*B*, *C* and S9). The predictions made by our computational model mirrored the results of aerotaxis experiments (Fig. 1*A*, 5*B* and *C*). To extend our model, we incorporated data on $O_2$-evoked changes in speed (Fig. 4*C* and S10), in addition to $O_2$-evoked changes in reversals. We found this did not substantially change the performance of *glb-5(tm5440)* and *glb-5(Haw)* in virtual aerotaxis assays. Model worms that modulated both reversals and speed in response to $O_2$ changes distributed similarly to animals that modulated only reversal (Fig. S11). By contrast, model worms in which

changes in $O_2$ influenced only speed distributed almost evenly in a virtual aerotaxis chamber (Fig. S11). Thus, in our model the relationship between URX response and reversal frequency is sufficient to account for the worm's $O_2$ preference in a shallow $O_2$ gradient.

**Discussion**

The neuroglobin GLB-5 changes how the URX $O_2$ sensing neurons encode $O_2$ concentration. URX sensory receptors enable *C. elegans* to avoid and escape 21% $O_2$. We find that URX neurons combine two putative molecular $O_2$ sensors, a soluble guanylate cyclase and a neuroglobin, to sculpt a sigmoidal $O_2$ tuning curve in which the neurons show little $Ca^{2+}$ response to stimuli below 13% $O_2$, gradually increase their responsiveness above this $O_2$ concentration, and begin to saturate as $O_2$ approaches 21%. The neuroglobin GLB-5 imposes the sigmoidal function by inhibiting the $O_2$-evoked $Ca^{2+}$ response in URX when $O_2$ levels fall below 21%. When GLB-5 is defective, the URX stimulus-response curve is shifted to lower $O_2$ levels and approaches saturation at 14% $O_2$. At a behavioral level, the effects of GLB-5 signaling is to broaden the $O_2$ environments preferred by *C. elegans* while maintaining strong avoidance of 21% $O_2$. If *glb-5* is defective, as in the N2 lab strain, animals prefer a narrow $O_2$ range, from 7% – 10%. Animals with functional *glb-5* signaling distribute more broadly, from 17% to 5% $O_2$. Some sensory responses exhibit steep sigmoidal tuning curves and it will be interesting to explore how frequently this is achieved by combining antagonistic molecular sensors. Studies of $O_2$ sensing in the glomus cells of the carotid bodies of mammals have implicated multiple $O_2$-sensing mechanisms that could act together to sculpt $O_2$ response features (Lopez-Barneo et al., 2016). Similarly, a range of $CO_2$/pH responsive molecules have been identified in mammals, although whether any of the numerous $CO_2$/pH-responsive cells use a combination of transducers is unclear (Huckstepp and Dale, 2011).

Unexpectedly, we find that the relationship between URX $Ca^{2+}$ response (a proxy of $O_2$ stimulus intensity) and behavioral output is nonlinear. Whereas intermediate stimulation of URX induces animals to reverse, strong stimulation is less effective. We have not investigated the neural mechanisms that underpin non-linear control of reversals by URX. However the neuroanatomical reconstructions reveal synapses from URX to both AVE interneurons that promote reversals, and to AVB interneurons that promote forward movement (White et al., 1986; Chalfie et al., 1985) (http://wormwiring.hpc.einstein.yu.edu/) which could be differentially regulated according to URX stimulation.

Several computational model have been constructed to elucidate behavioral mechanisms underlying *C. elegans* taxis behavior (Pierce-Shimomura et al., 1999; Iino and Yoshida, 2009; Roberts et al., 2016) These models have been built using detailed observation of animals moving in gradients. A taxis model that incorporates experimentally-measured neural activities has not, however, been reported, but is required to understand how neural signals are processed and transformed to behavior. We incorporated URX $Ca^{2+}$ responses measured using GCaMP6s into a random walk model. These data can be extended into a more detailed model e. g. incorporating activities of interneurons and motor neurons, to probe the relationship between neural coding in sensory neurons, downstream neural circuits, and behavioral output.

How does the GLB-5 neuroglobin alter the $Ca^{2+}$ responses of neurons at a molecular level? Like mammalian neuroglobin (Dewilde et al., 2001), GLB-5 rapidly oxidizes to a ferric form at 21% $O_2$ (Persson et al., 2009), suggesting it could participate in ROS (reactive oxygen species) or redox signaling. In URX GLB-5 co-localizes with GCY-35/GCY-36 soluble guanylate cyclases at dendritic endings(Gross et al., 2014; McGrath et al., 2009), and could potentially regulate the function of this other heme-binding protein. Our cGMP imaging did not support this hypothesis: we did not observe GLB-5-dependent differences in the $O_2$-evoked responses of URX. However, the cGMP dynamics we measured in the URX cell body were very slow compared to the $Ca^{2+}$ response, which implies that we are measuring a highly filtered response compared to the cGMP dynamics pertaining at the cGMP-gated channel. Although we do not exclude a role for GLB-5 in regulating soluble guanylate cyclases, our data suggest that GLB-5 can alter neural responses independently of these molecules. Expressing GLB-5 in the AFD sensory neurons altered their $CO_2$-evoked $Ca^{2+}$ responses; AFD neurons are not known to express soluble guanylate cyclases.

    Neuroglobin has been suggested to protect neurons from hypoxia (Burmester and Hankeln, 2009). Our study shows that a neuroglobin can participate in neural information processing. The *C. elegans* genome encodes a variety of other neurally expressed globins that may similarly modify neural function (Tilleman et al., 2011). It would be interesting to investigate whether neuroglobin participates in information processing in vertebrate neural circuits.

**Methods**

**Strains**

Strains were grown at 22-23 °C under standard conditions on Nematode Growth Medium (NGM) seeded with *Escherichia coli* OP50 (Sulston and Hodgkin, 1988).

**Neural imaging**

Immobilized animals: Animals expressing GCaMP6s, cGi500, or YC3.60 were glued to agarose pads (2% in M9 buffer) using Dermabond tissue adhesive (Ethicon) with the nose and tail immersed in *E. coli* OP50. The glued worms were then covered with a PDMS (polydimethylsiloxane) miculofluidic chamber, as described previously (Busch et al., 2012), and imaged using a 40× C-Apochromat lens on an inverted microscope (Axiovert; Zeiss) equipped with a Dual View emission splitter (Photometrics) and a Cascade II 512 EMCCD camera (Photometrics). The filters used were: GCaMP6s/mCherry: ex480/15 and 565/15 nm, di525/25 and 625/45 nm, em520/30 nm, em630/50 nm, di565 nm; YFP-CFP FRET: ex430/20 nm, di450 nm, em480/30 nm, em535/40 nm, di505 nm. Fluorescent images were captured at 1 fps with 2 x 2 or 1 x 1 binning using MetaMorph acquisition software (Molecular Devices). Data analysis used MATLAB (MathWorks) and Igor Pro (WaveMetrics). All time-lapse imaging data were denoised using binomial smoothing (Gaussian filter).

Delivery of gas stimuli: Humidified gas mixtures of defined composition were delivered using a PHD 2000 Infusion syringe pump (Harvard apparatus) at flow rates of 3.0, 2.0 and 1.0 ml/min. The syringes containing the gas were connected to PDMS chambers via polyethylene tubing and Teflon valves (AutoMate Scientific). A custom-built frame counter switched the valves at precise time points using TTL (transistor-transistor logic) pulses from the camera. To create the ramp stimulation, backlash air from the outlet of the PDMS chamber was used. The $O_2$ stimuli in chambers were measured using an $O_2$ probe (Oxygen Sensor Spots PSt3, PreSens).

**Behavioral assays**

Aerotaxis assays were performed as described previously (Gray et al., 2004) and animal positions noted 25 minutes into the assay. Briefly, rectangular PDMS chambers (dimensions of 33 x 15 x 0.2 mm) connected at either end to syringe pumps that delivered the indicated gas concentration were placed over 50 – 100 worms on a 9 cm NGM agar plate with food (*E. coli* OP50). The distribution of worms was recorded by counting animals in each of nine equal areas of the chamber.

To measure behavioral responses to step $O_2$-stimuli 5 adult hermaphrodites were placed on NGM plates seeded 36 – 40 h earlier with 20 $\mu$l of *E. coli* OP50 grown in 2× TY medium. To create a behavioral arena with a defined atmosphere, we placed a PDMS chamber (1 × 1 × 0.2 cm) on top of the worms, with inlets connected to a PHD 2000 Infusion syringe pump (Harvard apparatus), and delivered humidified gas mixtures of defined composition at a flow rate of 3.0 ml/min. Movies were captured at 2fps using FlyCapture software (Point Grey) on a Point Grey Grasshopper camera (Point Grey) mounted on a Leica M165FC stereo microscope. Movies were analysed using custom-written MATLAB software to calculate instantaneous speed. Instantaneous speed data were denoised by binning over 6 seconds. Reversal frequency was counted manually. If the posterior and anterior tips of a worm's body moved backward until the worm stopped, this behavior was counted as 1 reversal; such events were often followed by turns.

**Computational experiments and modelling**

In the computational model, a worm was represented as a single point in a virtual field that represented an $O_2$ gradient in our experimental 18 mm (W) x 15 mm (L) aerotaxis chamber. $O_2$ levels in the virtual chamber varied from 7 % at W=0 mm to 21 % at W = 18 mm. The worm moved forward either at constant (~0.05 mm/sec) or at variable speed. For iterations when speed varied according to $O_2$ concentration at the animal's position we acquired parameters for speed by performing curve fitting with a Hill equation using the speed data shown in Fig. 4*C* (Fig. S10*B* and *C*). The trend in an averaged time series was identified and removed based on gradient of the time series except frames from 241 to 300 (Fig. S10*A*). If the worm reached the edge of the chamber, the direction of forward movement was reflected. The model worm has three modules that correspond to the sensory neuron (URX), an interneuron, and a command neuron (Fig. 5*A*). The information signal about $O_2$ level is transmitted from the sensory neuron to the command neuron through the interneuron. The activation of the command neuron leads a worm to start reversing. Reversals are expressed as a change in the direction of locomotion in the model. The locomotion direction after the reversal was randomly chosen from a uniform distribution ($0.20e$ because experimentally measured reversals contain turning events. We assumed that the relationship between URX responses and reversal frequency was approximately linear in our model. This applies because animals in the virtual $O_2$ gradient, like those in a real-life aerotaxis assay, do not encounter large step $O_2$ stimuli.

The dynamics of the sensory circuit were represented by a Nonlinear-Linear-Nonlinear (NLN) model. The NLN model consisted of two nonlinear static filters and a linear temporal filter. $O_2$ stimulation was first converted by the input nonlinear filter, and processed by the temporal filter, then converted by the output nonlinear filter. The nonlinear filter $f(x)$ was expressed using a Hill equation,

$$f(x) = x^n/(x^n + x_0^n),$$

where $x_0$ and $n$ were the parameters that defined the range and strength of the nonlinearity of the filter, respectively. For convenience, the input and output nonlinear filter are hereafter denoted as $f_{in}$ and $f_{out}$, respectively. The linear temporal filter $K$ has a 361 sample length (t = $0,1, ... 360$) and satisfies

$$y = UK,$$

where $U$ and $y$ are the input and output time courses of the temporal filter, respectively. This typical expression of temporal filter should be expanded because our dataset has multiple time courses (multi dose). If O₂ concentration is left as $x$, $U$ can be written as

$$U = [U_1 \quad U_2 \quad ... \quad U_{dmax}]^T,$$

$$U_d = f_{in}\left(\begin{bmatrix} x_d(0-t_{max}) & x_d(1-t_{max}) & ... & x_d(0) \\ x_d(1-t_{max}) & x_d(2-t_{max}) & ... & x_d(1) \\ \vdots & \vdots & \ddots & \vdots \\ x_d(0) & x_d(1) & ... & x_d(t_{max}) \end{bmatrix}\right),$$

where $x_d(t)$ corresponds to the O₂ concentration at time $t$ of $d$-th step stimulation and $x_d(t < 0)$ is replaced by $x_d(0)$. $y$ can be expressed as

$$y = [y_1 \quad y_2 \quad ... \quad y_{dmax}]^T,$$

$$y_d = [y_d(0) \quad y_d(1) \quad ... \quad y_d(t_{max})]^T,$$

where $f_{out}(y_d(t))$ corresponds to the response of the sensory neuron at time $t$ in response to $d$-th oxygen step stimulation. The linear temporal filter $K$ can be obtained by evaluating

$$K = (U^T U)^{-1} U^T y.$$

For denoising, singular value decomposition was applied and the largest 100 components were used. In order to find the value of parameters of nonlinear filters, Nelder-Mead simplex optimization method was used and the sum of the square difference between $f_{out}(y_d(t))$ and corresponding experimental Ca²⁺ responses of URX were minimized. Because this optimization was done separately for *glb-5(Haw)* and *glb-5(tm5440)*, we obtained two parameter sets for the NLN model.

      Interneuron and command neuron were reasonably designed as described below. Because the worms show random reversals, the command neuron should be randomly activated. Furthermore, because the basal URX Ca²⁺ response (i.e. before stimulation of 2% change of oxygen) depends on the basal concentration of O₂ but basal

reversal frequency does not, the model should contain a temporal differentiation functionality. Therefore, the activity of interneuron $g(t)$ was modelled as

$$g(t) = f_{out}(y_d(t)) - \sum_{\tau=1}^{l} f_{out}(y_d(t-\tau))/l ,$$

where $l$ is a lag constant and was fixed as 11. The activity of command neuron is positive when

$$r < b * (1 + g(t) * c)$$

where $b$ is the basal reversal frequency that is computed from experimental data, $c$ is the coefficient of the effect of $O_2$ stimulation, and $r$ is a uniformly distributed random number between 0 and 1. $b$ and $c$ were fixed to 0.0723 (reversal frequency per 1 second before a stimulation is given) and 3, respectively. Note that the parameters of interneuron and command neuron ($l$, $b$, and $c$) are independent of the *glb-5* genotype.

**Strain list**

AX5890, *glb-5(tm5440); npr-1(ad609); dbEx[gcy-32p::GCaMP6s, gcy-32p::mCherry, unc-122p::mCherry]*

AX5891, *glb-5(Haw); npr-1(ad609); dbEx[gcy-32p::GCaMP6s, gcy-32p::mCherry, unc-122p::mCherry]*

AX6075 *gcy-35(ok769); glb-5(tm5440); npr-1(ad609); dbEx[gcy-32p::GCaMP6s, gcy-32p::mCherry, unc-122p::mCherry]]*

AX6088, *glb-5(tm5440); npr-1(ad609); dbEx[gcy-32p::GCaMP6s, gcy-32p::mCherry, unc-122p::mCherry]; dbEx[gcy-32p::glb-5 (Haw)::sl-2::mCherry, unc-122p::GFP]*

AX1891, *glb-5(Haw); npr-1(ad609)*

AX5935, *gcy-35(ok769); glb-5(tm5440); npr-1(ad609)*

AX5936, *gcy-35(ok769); glb-5(tm5440); npr-1(ad609); dbEx[flp-8p::gcy-35::gfp, unc-122p::gfp]*

AX6124, *gcy-35(ok769); glb-5(tm5440); npr-1(ad609); dbEx[flp-8p::gcy-35::gfp, unc-122p::gfp], dbEx[flp-8p::glb-5::sl-2::mCherry, lin-44p::GFP]*

AX1908, *glb-5(Haw); npr-1(ad609) lin-15(n765ts); dbEx[gcy-32p::YC3.60, lin-15(+)]*

AX5850, *glb-5(tm5440); npr-1(ad609) lin-15(n765ts); dbEx[gcy-32p::YC3.60, lin-15(+)]*

AX3535, *glb-5(tm5440); npr-1(ad609)*

AX5779, *dbEx[gcy-8p::YC3.60, odr-1p::mCherry]; dbEx[gcy-8p::glb-5 (Haw)]*

AX2047, *dbEx[gcy-8p::YC3.60, odr-1p::mCherry]*

AX2417, *glb-5(Haw); cng-1(db111); npr-1(ad609); dbEx[gcy-37p::cGi500]*

AX6024, *glb-5(tm5440); cng-1(db111); npr-1(ad609); dbEx[gcy-37p::cGi500]*

AX2084, *glb-5(Haw); npr-1(ad609); dbEx[gcy-37p::cGi500]*

AX6012, *glb-5(tm5440); npr-1(ad609); dbEx[gcy-37p::cGi500]*

**Author Contributions**: S.O. designed research, performed experiments and analysed experimental data. All authors interpreted data and wrote the paper. Y.T. and S.O. constructed computational models, performed computational experiments and analysed results of computational experiments.

**Acknowledgements**

We thank the National Bioresource Project for strains, W.R. Schafer and Y. Iino for plasmids, and I. Beets, L. Fenk, T. Tomida and S. Laughlin for comments on the manuscript. This work was supported by the Medical Research Council (MC_U105178786), European Research Council (269058 ACMO), Uehara Memorial Foundation (S.O.), Grants-in-Aid for Young Scientists (B) (26830006) (Y.T.) and for Grant-in-Aid for Scientific Research on Innovative Areas (16H01418) from the Ministry of Education, Culture, Sports, Science and Technology of Japan (Y.T.).

**References**

Bretscher, A. J., Kodama-Namba, E., Busch, K. E., Murphy, R. J., Soltesz, Z., Laurent, P., and de Bono, M. (2011). Temperature, Oxygen, and Salt-Sensing Neurons in C. elegans Are Carbon Dioxide Sensors that Control Avoidance Behavior. Neuron *69*, 1099-1113.

Burmester, T., and Hankeln, T. (2009). What is the function of neuroglobin? J Exp Biol *212*, 1423-1428.

Burmester, T., and Hankeln, T. (2014). Function and evolution of vertebrate globins. Acta Physiol (Oxf) *211*, 501-514.

Burmester, T., Weich, B., Reinhardt, S., and Hankeln, T. (2000). A vertebrate globin expressed in the brain. Nature *407*, 520-523.

Busch, K. E., Laurent, P., Soltesz, Z., Murphy, R. J., Faivre, O., Hedwig, B., Thomas, M., Smith, H. L., and de Bono, M. (2012). Tonic signaling from $O_2$ sensors sets neural circuit activity and behavioral state. Nat Neurosci *15*, 581-591.

Chalfie, M., Sulston, J. E., White, J. G., Southgate, E., Thomson, J. N., and Brenner, S. (1985). The neural circuit for touch sensitivity in *Caenorhabditis elegans*. J Neurosci *5*, 956-964.

Chen, T. W., Wardill, T. J., Sun, Y., Pulver, S. R., Renninger, S. L., Baohan, A., Schreiter, E. R., Kerr, R. A., Orger, M. B., Jayaraman, V., Looger, L. L., Svoboda, K., and Kim, D. S. (2013). Ultrasensitive fluorescent proteins for imaging neuronal activity. Nature *499*, 295-300.
19

**Figure Legends**

**Fig. 1. The GLB-5 neuroglobin broadens *C. elegans*' $O_2$ preference**

(A) Aerotaxis behavior. Distribution of animals in a 0 – 21% $O_2$ gradient. N = 6. Plots show mean +/- s.e.m. (B) The *glb-5(Haw)* allele broadens *C. elegans'* $O_2$ preference. High $O_2$ avoidance index = (fraction of animals in 7%–14% $O_2$) – (fraction of animals in 14–21% $O_2$) / (fraction of animals in 7– 21% $O_2$). $O_2$ preference (7–10 % $O_2$) = (fraction of animals in 7%–10% $O_2$) / (fraction of animals in 7%– 21% $O_2$). **$p < 0.01$, Mann-Whitney U-test. Data are from (A). (C) Mean $Ca^{2+}$ responses of URX and PQR neurons to the indicated $O_2$ ramp stimulus. N = 10–12. Shading represents s.e.m. $Ca^{2+}$ sensors were GCaMP6s co-expressed with mCherry (URX) and YC3.60 (PQR). The URX plot also shows the responses of *gcy-35; glb-5(tm5440)* mutants; GCY-35 is required for measurable $O_2$-evoked $Ca^{2+}$ responses. The $O_2$ ramp stimulus (top) shows the mean of 4 measurements +/- s.e.m.

**Fig. 2. GLB-5 changes the stimulus-response curve of $O_2$-evoked $Ca^{2+}$ responses in URX**

(A) $Ca^{2+}$ responses evoked in URX by a graded series of $O_2$ steps that start from a 7% $O_2$ baseline (N = 10). Data show mean +/- s.e.m. Responses are normalised to the GCaMP6s/mCherry ratio averaged over the 10 secs prior to delivery of the stimulus train. (B) Maximum amplitude of the responses from (A) plotted against $O_2$ stimulus intensity. Bars represent s.e.m. *, $p < 0.05$; ** $p < 0.01$; *glb-5(Haw)* vs *glb-5(tm5440)*. ++, $p < 0.01$; *glb-5(tm5440) URXp::glb-5(Haw)* vs *glb-5(tm5440)*. ANOVA with Dunnett's post hoc test.

**Fig. 3. Coding of step and ramp $O_2$ stimuli by URX**

(A) Averaged traces of the $Ca^{2+}$ response of URX to different 2% $O_2$ step stimuli in *glb-5(Haw)*, *glb-5(tm5440)*, and *glb-5(tm5440); pgcy-32::glb-5(Haw)* animals. N = 10-18. (B), Maximum amplitudes of the responses shown in (A) plotted against stimulus intensity. Data show mean +/- s.e.m.. (C) Averaged traces of the $Ca^{2+}$ response of URX to different 2% $O_2$ ramp stimuli in *glb-5(Haw)* and *glb-5(tm5440)* animals. N = 11 – 16. (D), Maximum amplitude of the responses shown in (C) plotted against stimulus intensity. Data show mean and s.e.m. The $O_2$ plots of a step and a ramp stimulus at top of (A) and (C) show the mean of 9 (A) and 10 (C) measurements +/- s.e.m. *, $p < 0.05$; ** $p < 0.01$; *glb-5(Haw)* vs *glb-5(tm5440)*. +, $p < 0.05$; ++, $p < 0.01$; *glb-5(tm5440) URXp::glb-5(Haw)* vs *glb-5(tm5440)*. ANOVA with Dunnett's post hoc test (B). Mann-Whitney U test (D).

**Fig. 4. The relationship between $O_2$ stimulus intensity and URX-dependent behavioral outputs**

(A) Frequency of reversal behavior evoked by $O_2$ step stimuli in animals of the genotypes indicated. Reversal frequencies were quantified every minute. N = 15-30 animals. Bars represent s.e.m. (B) Reversal frequency evoked by step $O_2$ stimuli averaged over 4 minutes after the stimulus. Data represent mean +/- s.e.m. (C and D) Instantaneous speed in response to step $O_2$ stimuli indicated. N = 30-60 animals. Plots show mean +/- s.e.m. (D) Mean speed +/- s.e.m calculated for a 4 minute interval beginning 30 seconds after the step stimulus. The behavior of *gcy-35(ok769); glb-5(tm5440)* animals was used as a negative control, and is shown as black traces or grey bars. *$P < 0.05$, **$P < 0.01$, ANOVA with Dunnett's post hoc test. NS, not significant.

**Fig. 5. A computational model that links $O_2$-evoked $Ca^{2+}$ responses in URX to behavioral output**

(A) Schematic of the computational model. (B) Heat map representing the location of 10000 fictive *glb-5(tm5440)* or *glb-5(Haw)* animals in a 7% to 21% $O_2$ gradient. Locations are plotted every second. (C) Histograms of the existence frequency of *glb-5(tm5440)* and *glb-5(Haw)* in (A) 7% to 21% $O_2$ gradient during the last 100 seconds of the computational experiments shown in (B). The fictive URX responses and reversal frequency of these model worms during aerotaxis are shown in Fig. S9.

**Supplementary Figure Legends**

**Fig. S1. The URX Ca$^{2+}$ response to strong stimulation**

(A) Mean URX Ca$^{2+}$ responses to a 7% to 21% $O_2$ stimulus in *glb-5(Haw)* (red trace) and *glb-5(tm5440)* (blue trace) animals. N = 8. Shading represents s.e.m. (B) Peak $\Delta R/R_0$ +/- s.e.m. evoked by the $O_2$ stimulus in (A). NS, not significant (Mann-Whitney U-test).

**Fig. S2. URX Ca$^{2+}$ responses to an exponential ramp $O_2$ stimulus in the absence of food**

Ca$^{2+}$ responses in URX evoked by an exponential ramp $O_2$ stimulus in *glb-5(tm5440)* (blue trace) and *glb-5(Haw)* (red trace) animals imaged in the absence of food. N = 10. Shading represents s.e.m. The Ca$^{2+}$ sensor was GCaMP6s (co-expressed with mCherry). The $O_2$ stimulus used is shown above the traces, plotted as the mean $O_2$ concentration +/- s.e.m. N = 4.

**Fig. S3. Expressing *glb-5(Haw)* in URX is sufficient to change *C. elegans'* $O_2$ preference**

(A) Aerotaxis of *C. elegans* expressing *gcy-35* or *gcy-35* and *glb-5(Haw)* selectively in URX. *glb-5(tm5440)* is a positive control and the same *dbExpflp-8::gcy-35* transgenic line was used with or without the *pURX::glb-5(Haw)* transgene. N = 10-12. Data plotted are Mean +/- s.e.m. (B and C) High $O_2$ avoidance index = (fraction of animals in 7%–14% $O_2$) – (fraction of animals in 14–21% $O_2$) / (fraction of animals in 7– 21% $O_2$). $O_2$ preference (7–10 % $O_2$) = (fraction of animals in 7%–10% $O_2$) / (fraction of animals in 7%– 21% $O_2$). *$p$ < 0.05, **$p$ < 0.01, ANOVA with Tukey's post hoc test. Data are from (A)

**Fig S4. $O_2$-evoked cGMP responses in URX**

(A) Model of sensory transduction of $O_2$ stimuli in URX. (B) cGMP responses evoked in URX by an exponential $O_2$ ramp stimulus. Data show Mean +/- s.e.m. Top: traces from *glb-5(tm5440)* (blue) and *glb-5(Haw)* (red) animals. Bottom: traces from *cng-1(db111); glb-5(tm5440)* (blue) and *cng-1(db111); glb-5(Haw)* (red) animals. N = 10-11. The $O_2$ stimulus used is plotted on top of the cGMP traces. Data show Mean $O_2$ +/- s.e.m. N = 4. (C) Average responses compared at time points indicated by the black bar in (B) (bottom). NS, not significant (Mann-Whitney U-test).

**Fig. S5. Ectopic expression of GLB-5(Haw) in AFD alters $CO_2$-evoked $Ca^{2+}$ responses in this neuron**

(A) Mean $Ca^{2+}$ responses evoked by a series of $CO_2$ stimuli in AFD neurons in animals with or without ectopic expression of *glb-5(Haw)* in AFD. N = 10-11. The sensor is YC3.60. The background $O_2$ concentration in these experiments was 11 % $O_2$. Traces were normalized using the average YFP/CFP ratio for the 10 secs immediately before delivery of each $CO_2$ step in the train. (B and C) Minimum (B) or maximum (C) values of AFD $Ca^{2+}$ responses upon $CO_2$ exposure or removal plotted against stimulation intensity. Values are normalized using the average YFP/CFP ratio for 10 the sec immediately prior to each $CO_2$ upstep (B) or downstep (C). Error bars represent s.e.m. *$P < 0.05$, Mann-Whitney U-test.

**Fig. S6. GLB-5(Haw) confers $O_2$ responsiveness to the AFD sensory neuron**

(A and B) Ca$^{2+}$ responses evoked by indicated O$_2$ stimuli in wild type AFD or AFD that ectopically express *glb-5(Haw).* N = 13-15. The sensor is YC3.60. (C and D) The maximum Ca$^{2+}$ response over a 30 second interval at 21% O$_2$ from a, b. * $p < 0.05$; ** $p < 0.01$, Mann-Whitney U-test.

**Fig. S7. Strong stimulation of URX does not induce reversals**

(A) Mean URX Ca$^{2+}$ response evoked by a 13 → 21% step O$_2$ stimulus in genotypes indicated. Shading represents s.e.m (N = 8–11). (B) Reversal frequencies of *C. elegans* in response to a 13 → 21 % step O$_2$ stimulus. Reversal frequency was quantified every one minute (N = 14–20 animals). Bars represent s.e.m.

**Fig. S8. NLN model: reconstruction of URX responses**

Reconstruction by the NLN model of Ca$^{2+}$ responses evoked in URX by 2% O$_2$ step stimuli in *glb-5(Haw)* and *glb-5(tm5440)* animals. Solid and dashed lines indicate reconstructed data and experimental data, respectively. The experimental data used for comparison are taken from Fig. 3*A*.

**Fig. S9. URX responses and reversal frequency of model worms in Fig. 5**

(A and B) Histograms of fictive URX responses of *glb-5(tm5440)* (A) and *glb-5(Haw)* (B) model worms over the 1800 sec of the *in silico* experiment shown in Fig. 4. (C) Simulated tracks of three worms navigating the O$_2$ gradient in our computer simulations. Shown are 1801 steps. The X and Y axis are the dimensions of the fictive aerotaxis chambers. Colors code local O$_2$ concentration**.** (D) Reversal frequency at each O$_2$ concentration of the model worms during the 1800 sec. Red and blue traces indicate fictive *glb-5(Haw)* and *glb-5(tm5440)* animals respectively.

**Fig. S10. Data processing of speed data for computational experiments**

(A) De-trending speed data. To incorporate experimental speed data into our model, we identified and removed a slow downward trend in an averaged time series of the speed of animals responding to $O_2$ cues. The raw experimental speed data (from Fig. 4C) are shown on the left, with the dashed lines showing the fitted lines used for detrending. The de-trended data is on the right. (B and C) De-trended average speed before (triangle) or after (circle) we delivered the $O_2$ stimuli described in Fig. 3. By performing curve fitting on this data we acquired parameters for the Hill equation used in the computational experiments.

**Fig. S11. Modeling how $O_2$-evoked changes in speed alters $O_2$-preference in aerotaxis assays**

Instantaneous location of 10,000 fictive *glb-5(tm5440)* and *glb-5(Haw)* animals in different aerotaxis models, plotted at 1 sec intervals and represented by heat maps. Top: fictive animals have constant speed but show $O_2$-evoked reversals according to our model for URX output. Middle: Fictive animals show $O_2$-evoked changes in speed according to our model for URX output but not in reversals. Bottom: fictive animals show $O_2$-evoked changes in both reversals and speed. Histograms show the existence frequency of model worms during the last 100 seconds of the computational experiments.

Figure 1

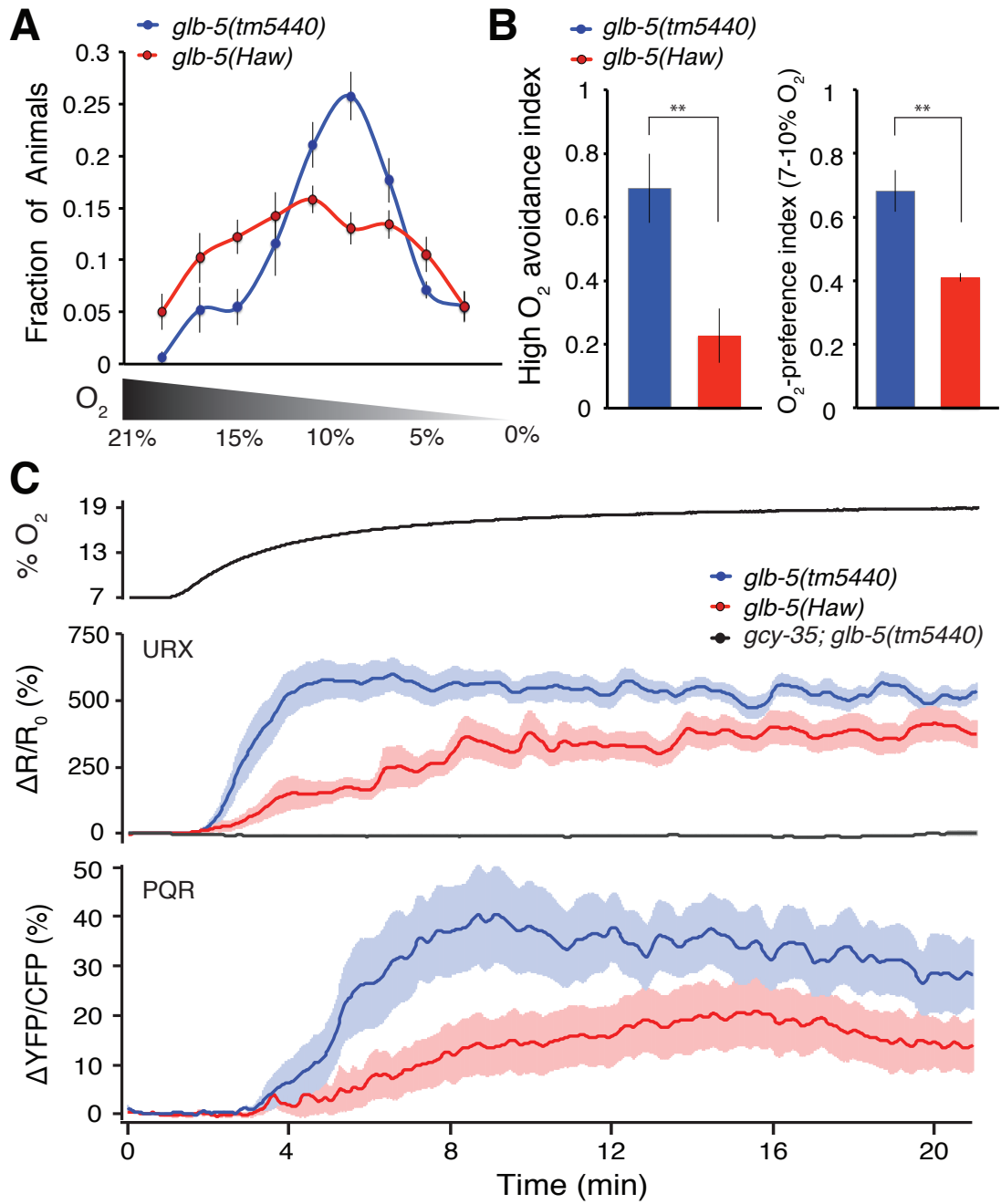

Figure 2

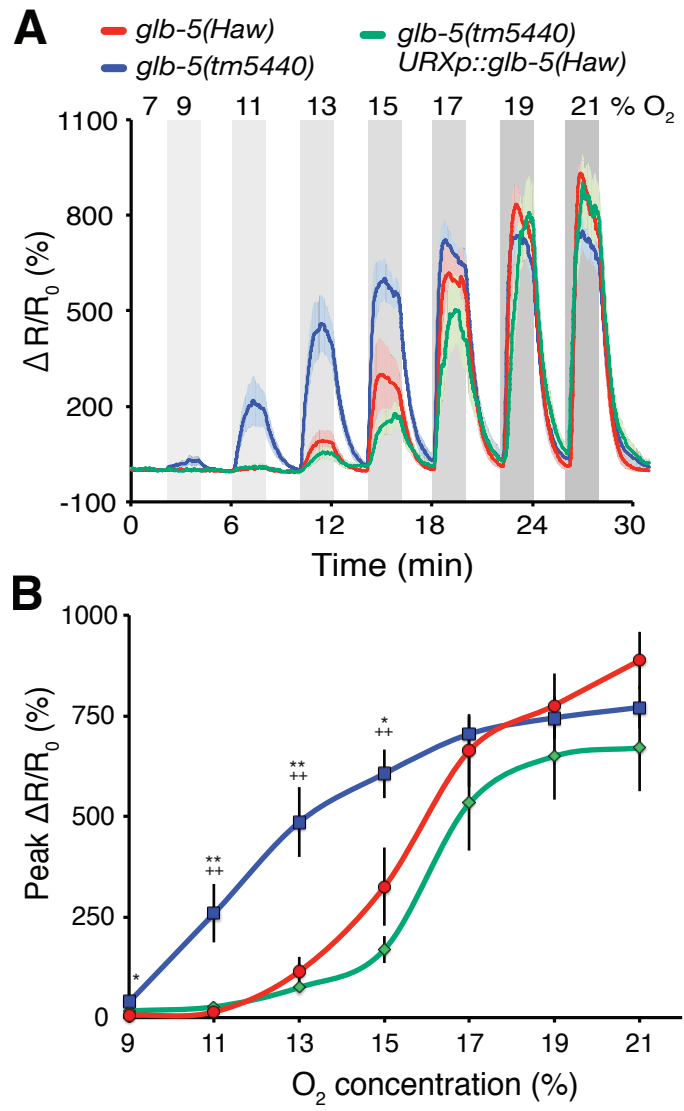

Figure 3

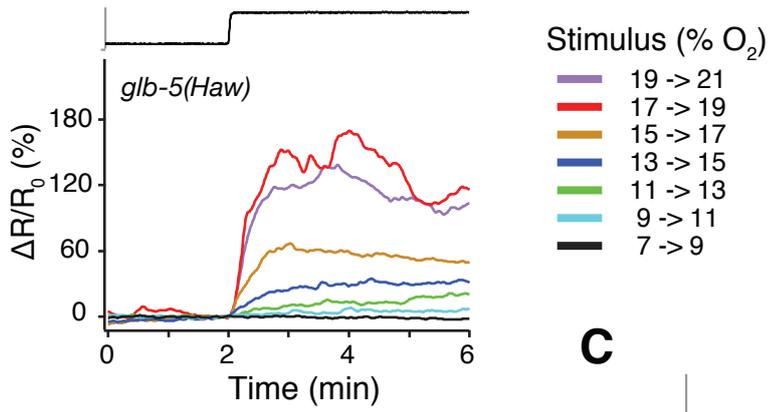
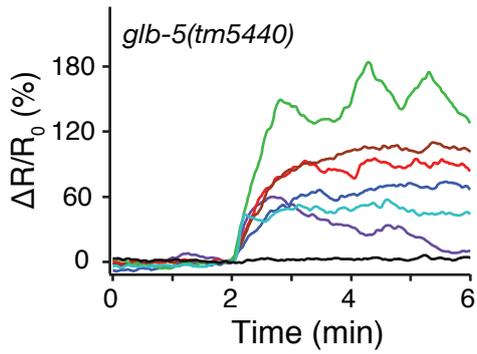
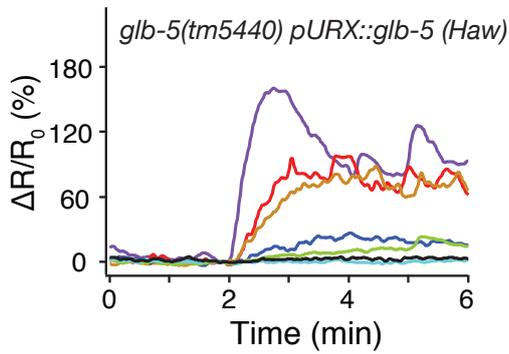
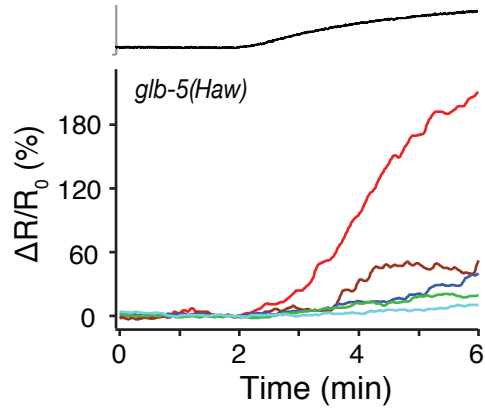
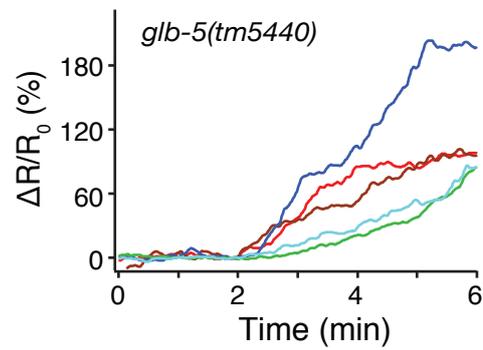
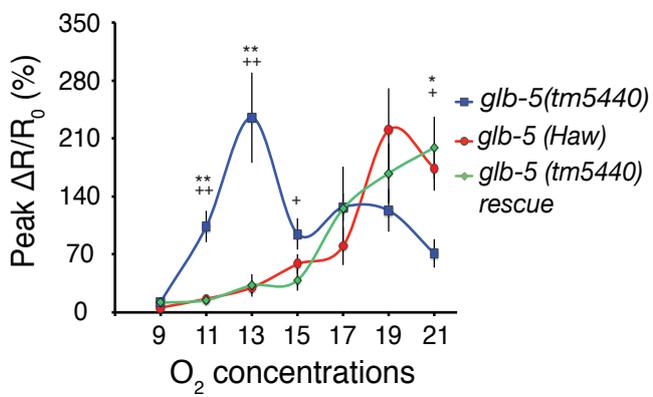
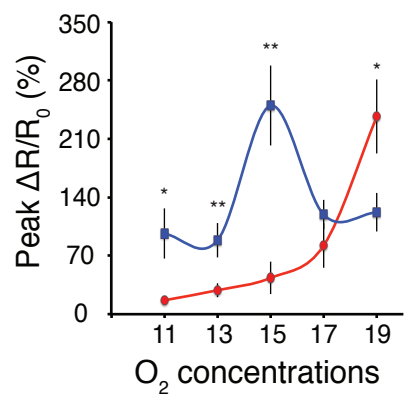

Figure 4

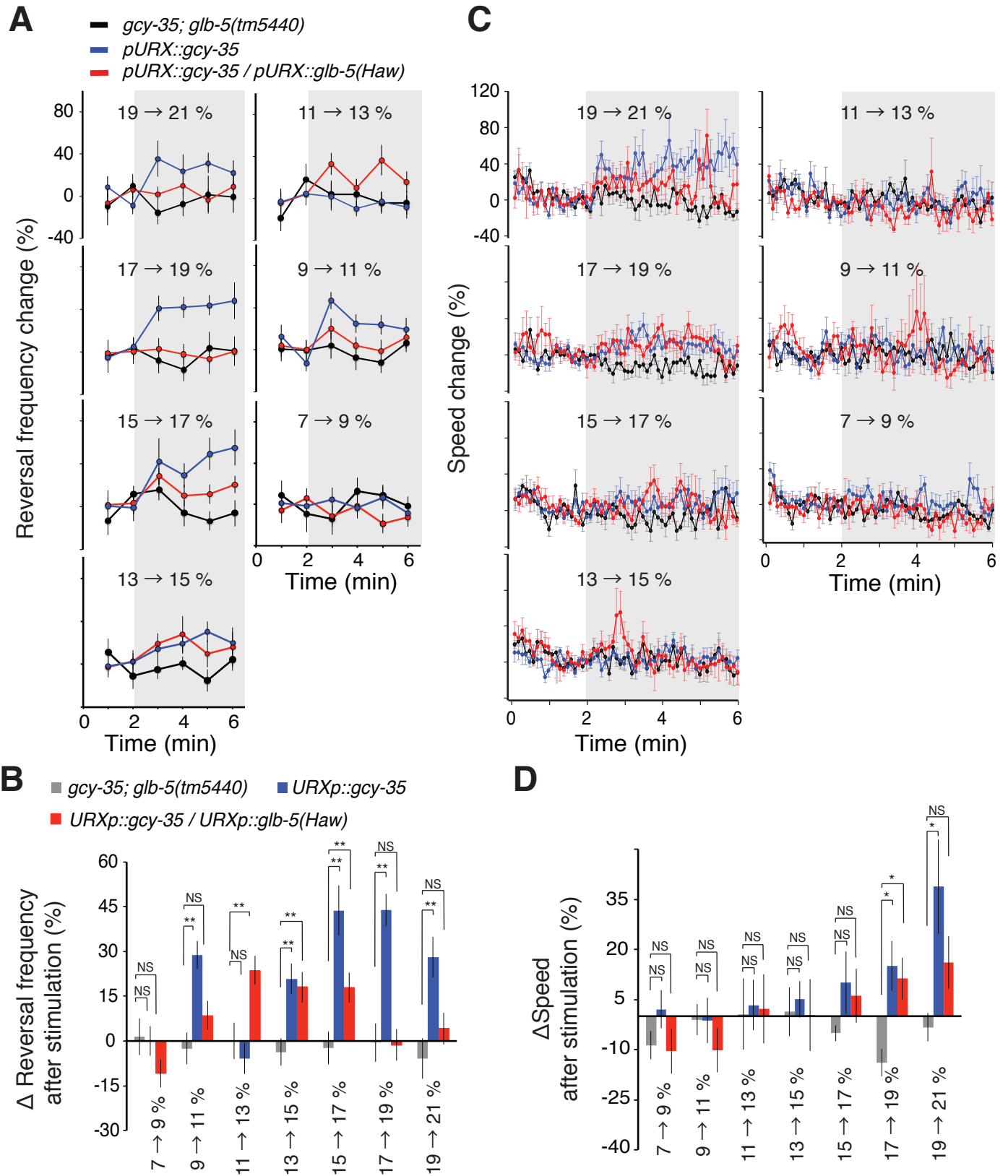

# Figure 5

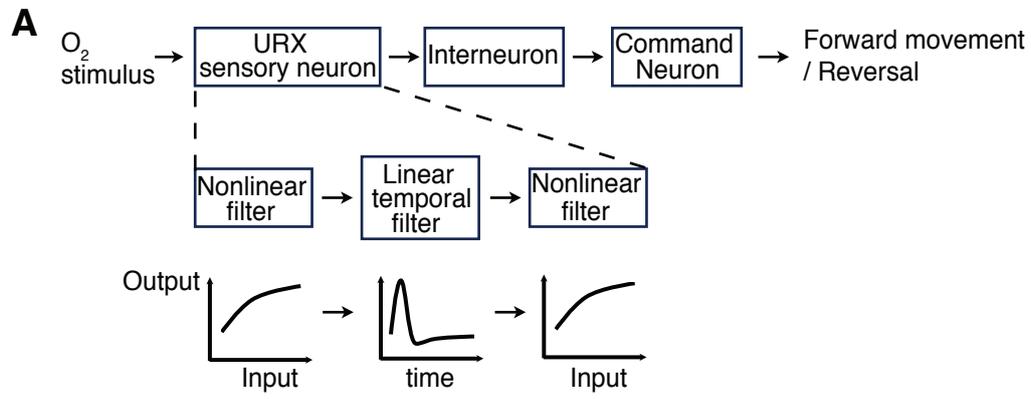
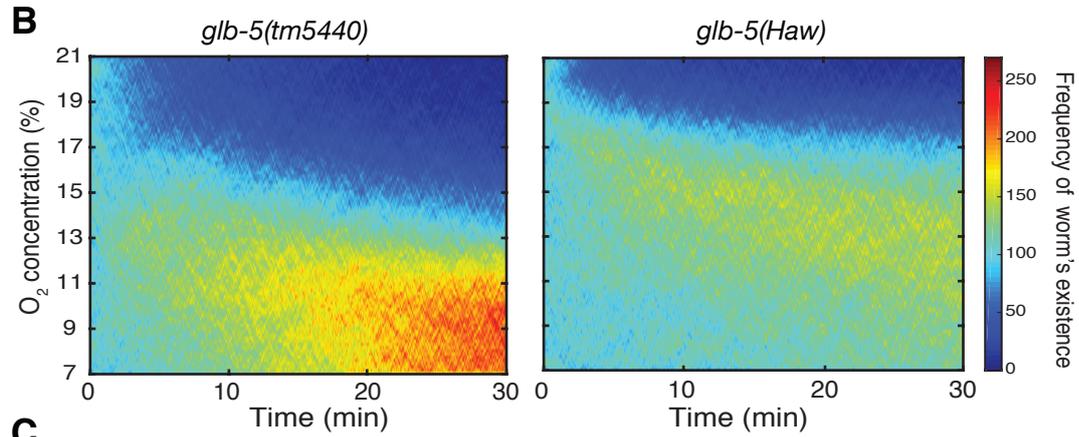
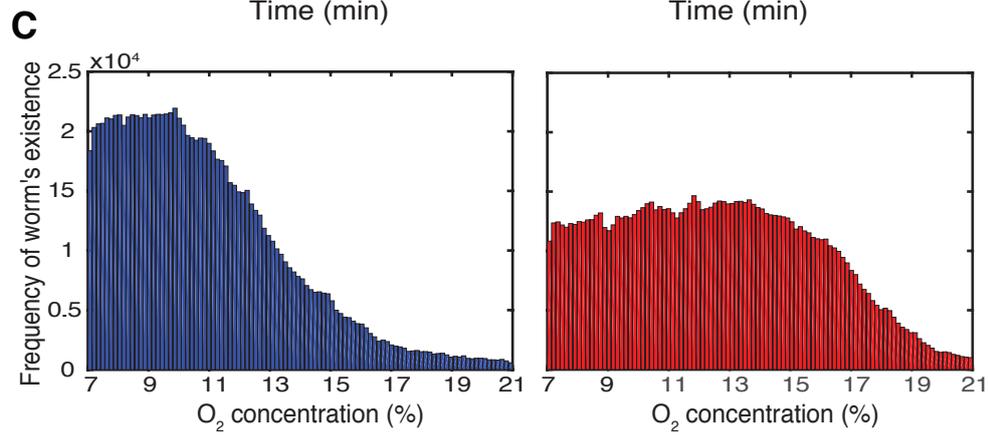

Fig S1

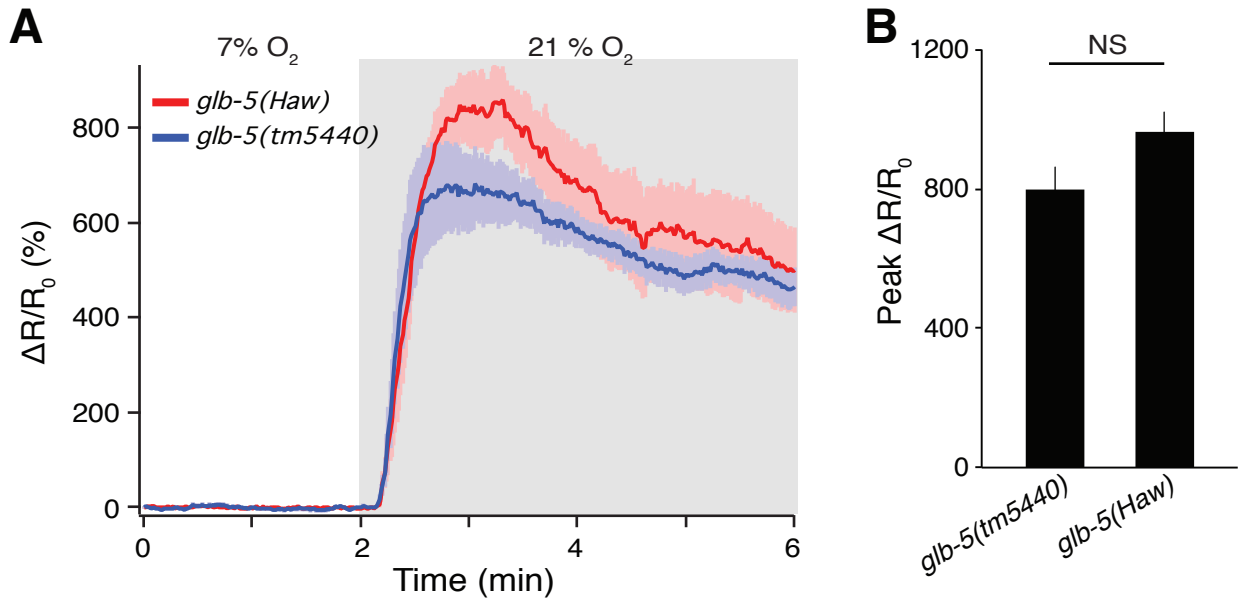

Fig S2

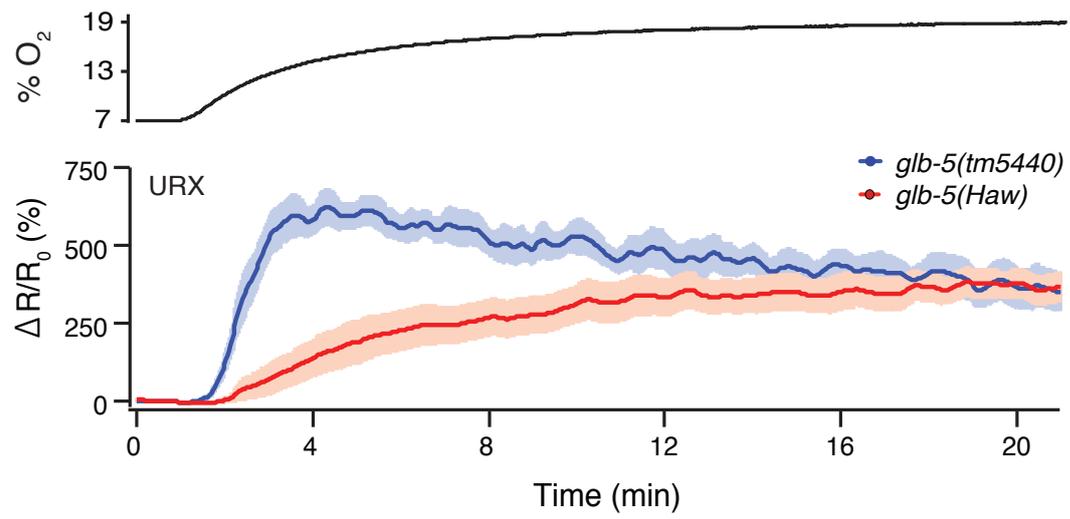

Fig S3

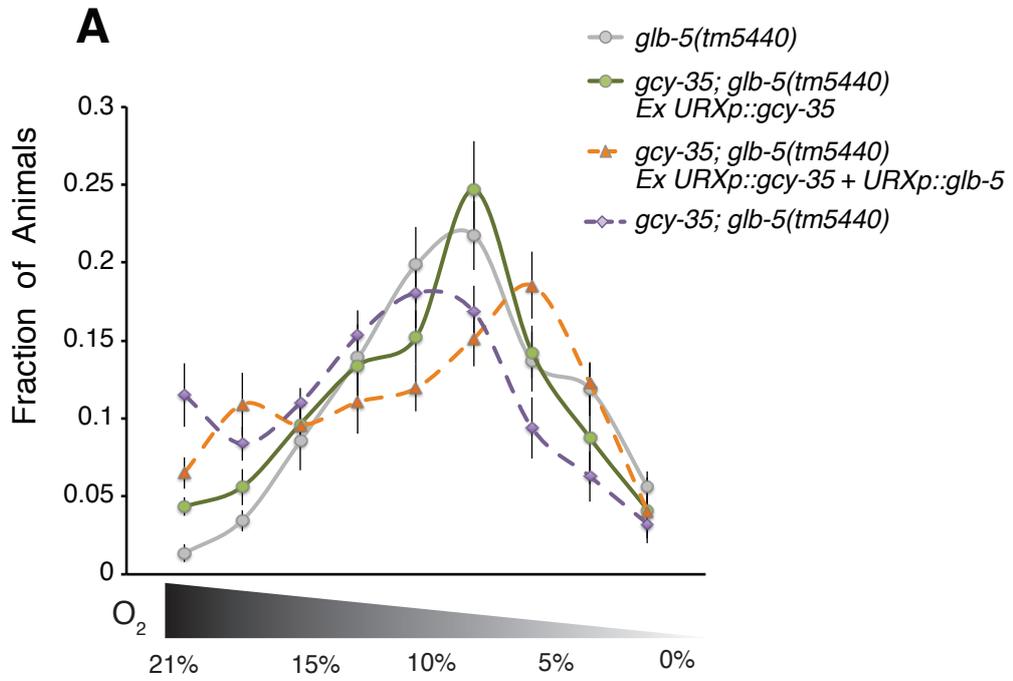

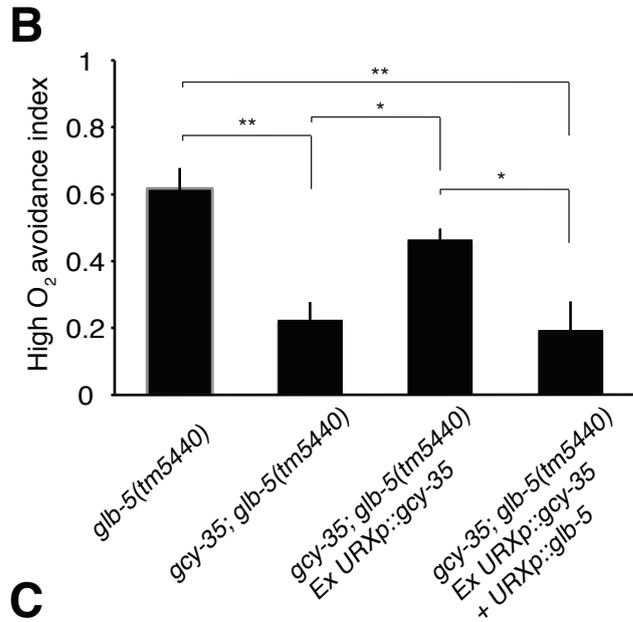

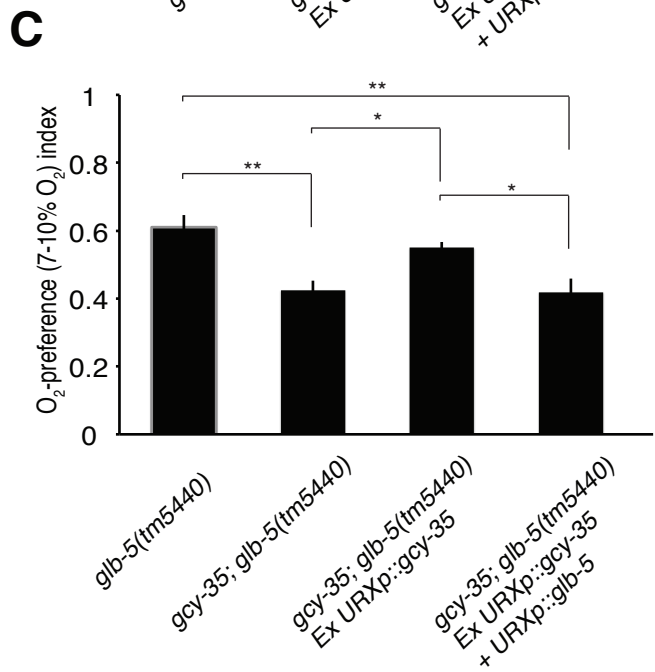

Fig S4

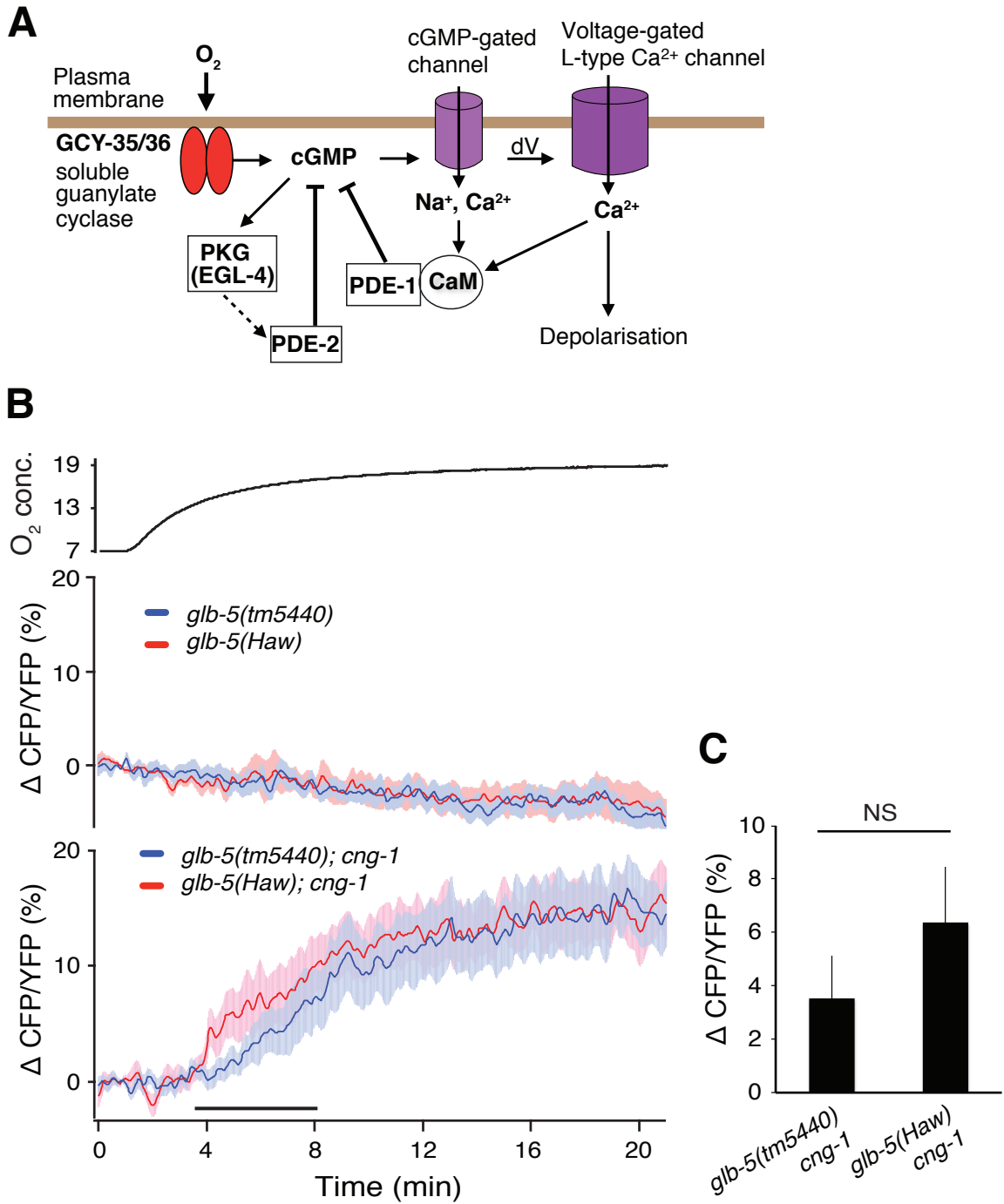

Fig S5

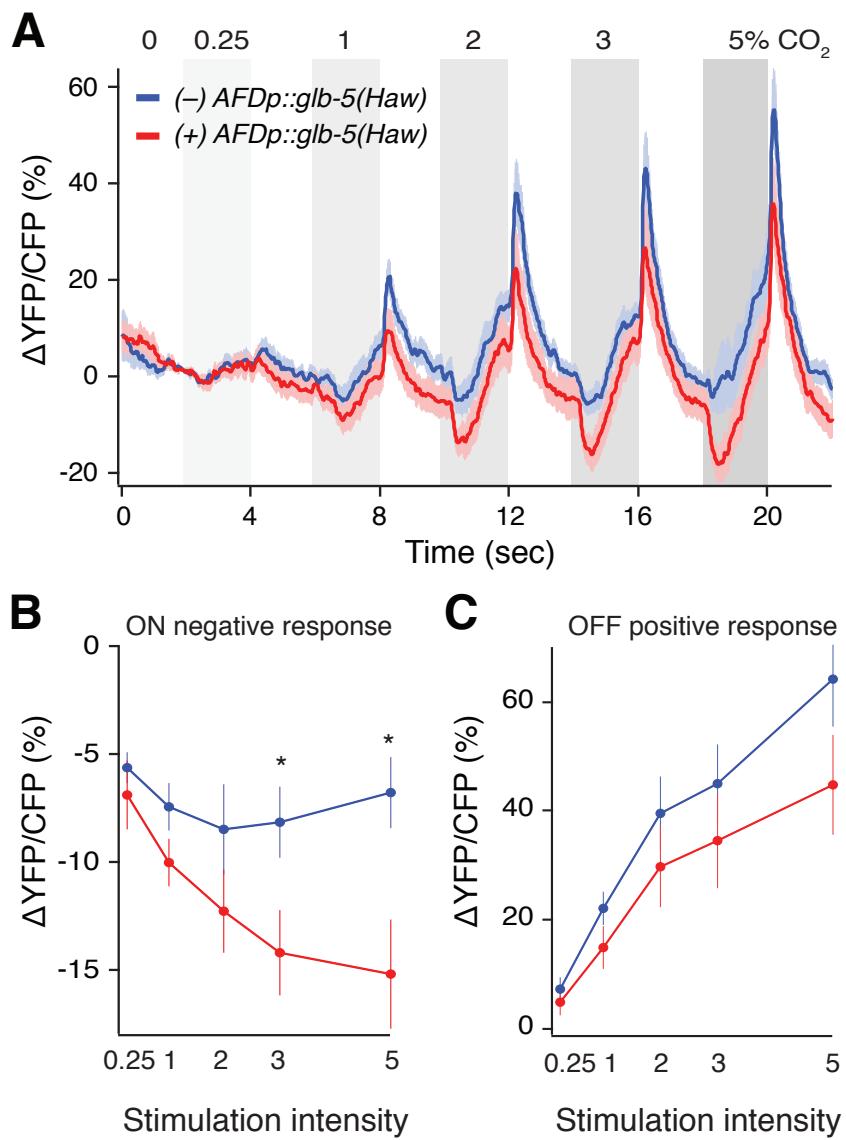

Fig S6

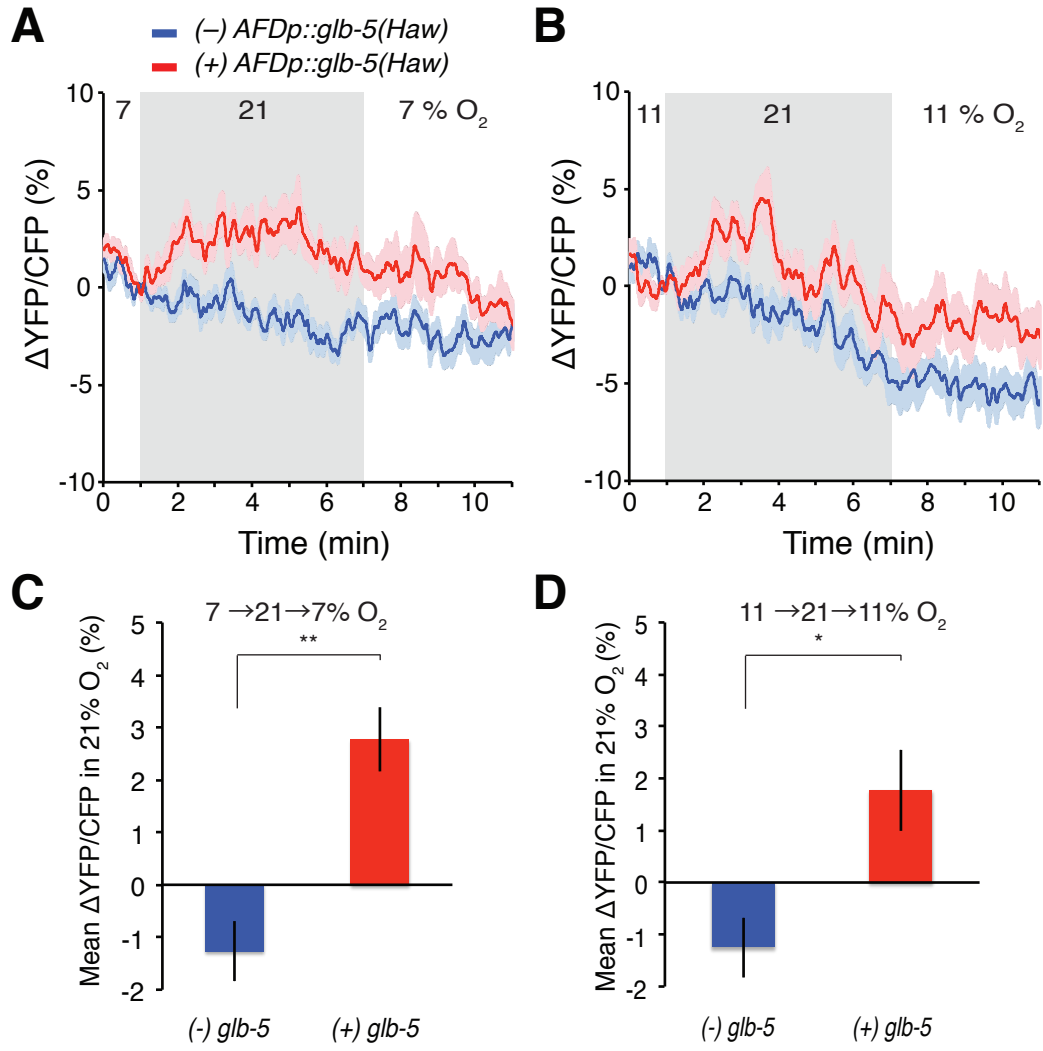

Fig S7

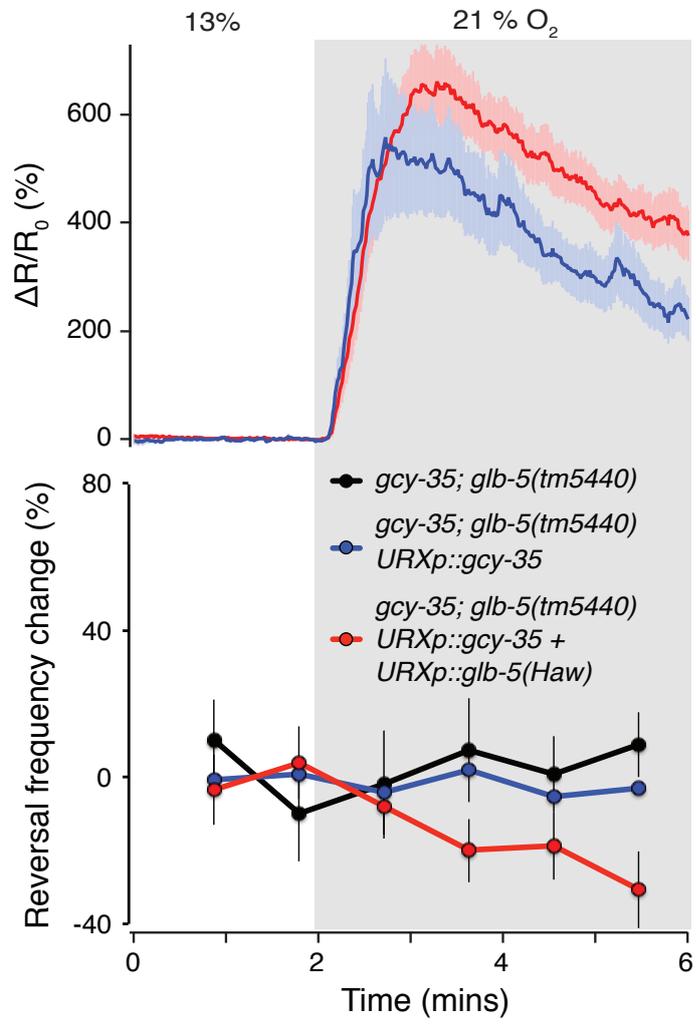

Fig S8

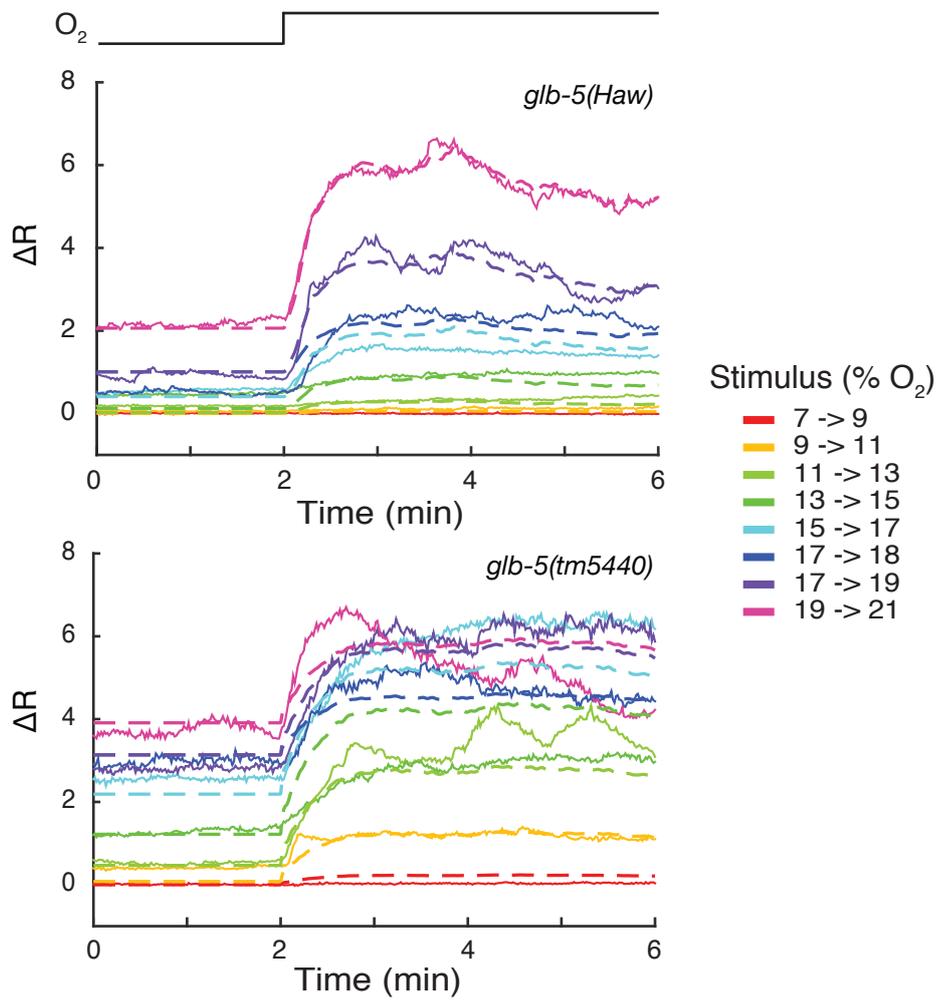

Fig S9

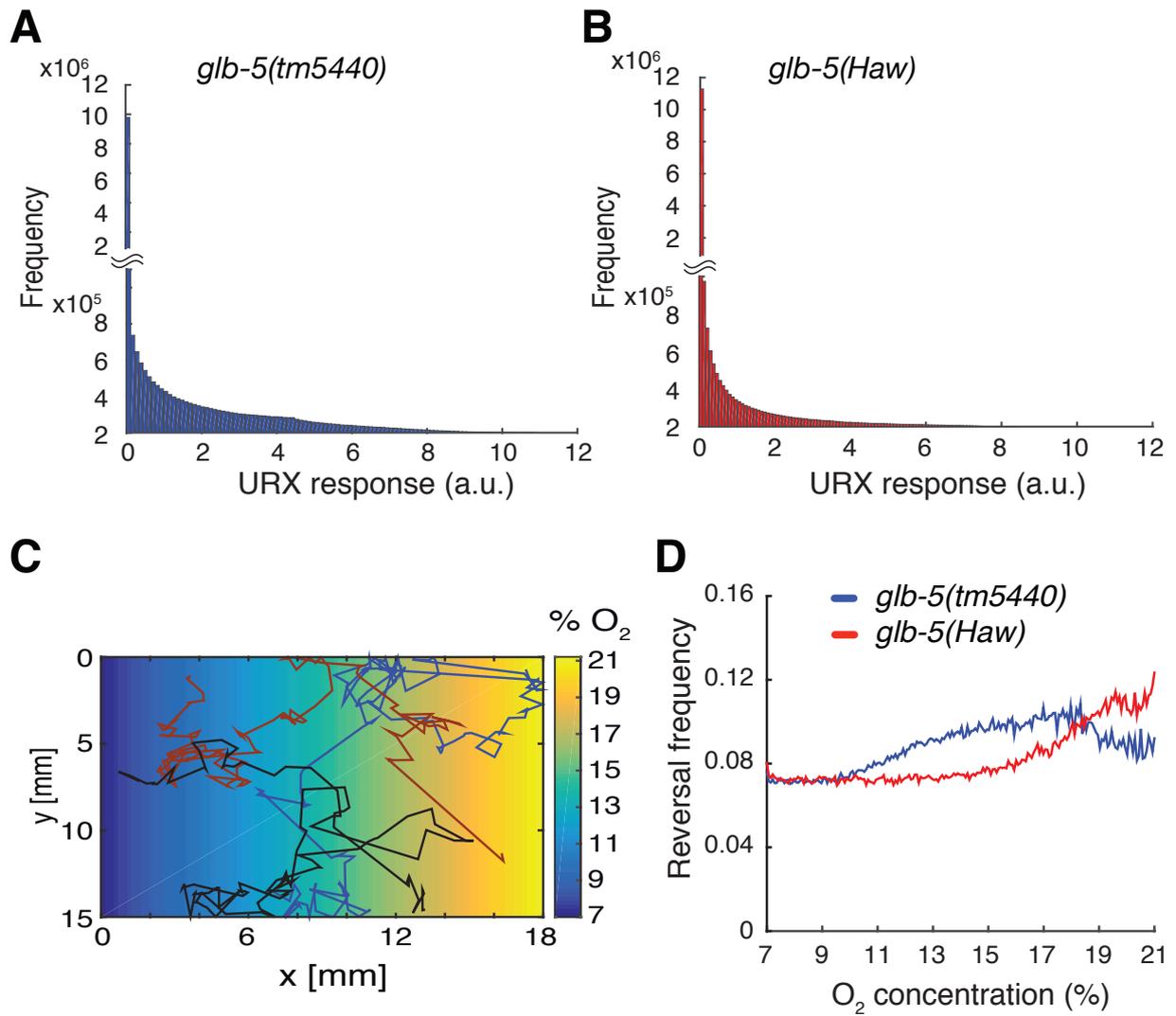

Fig S10

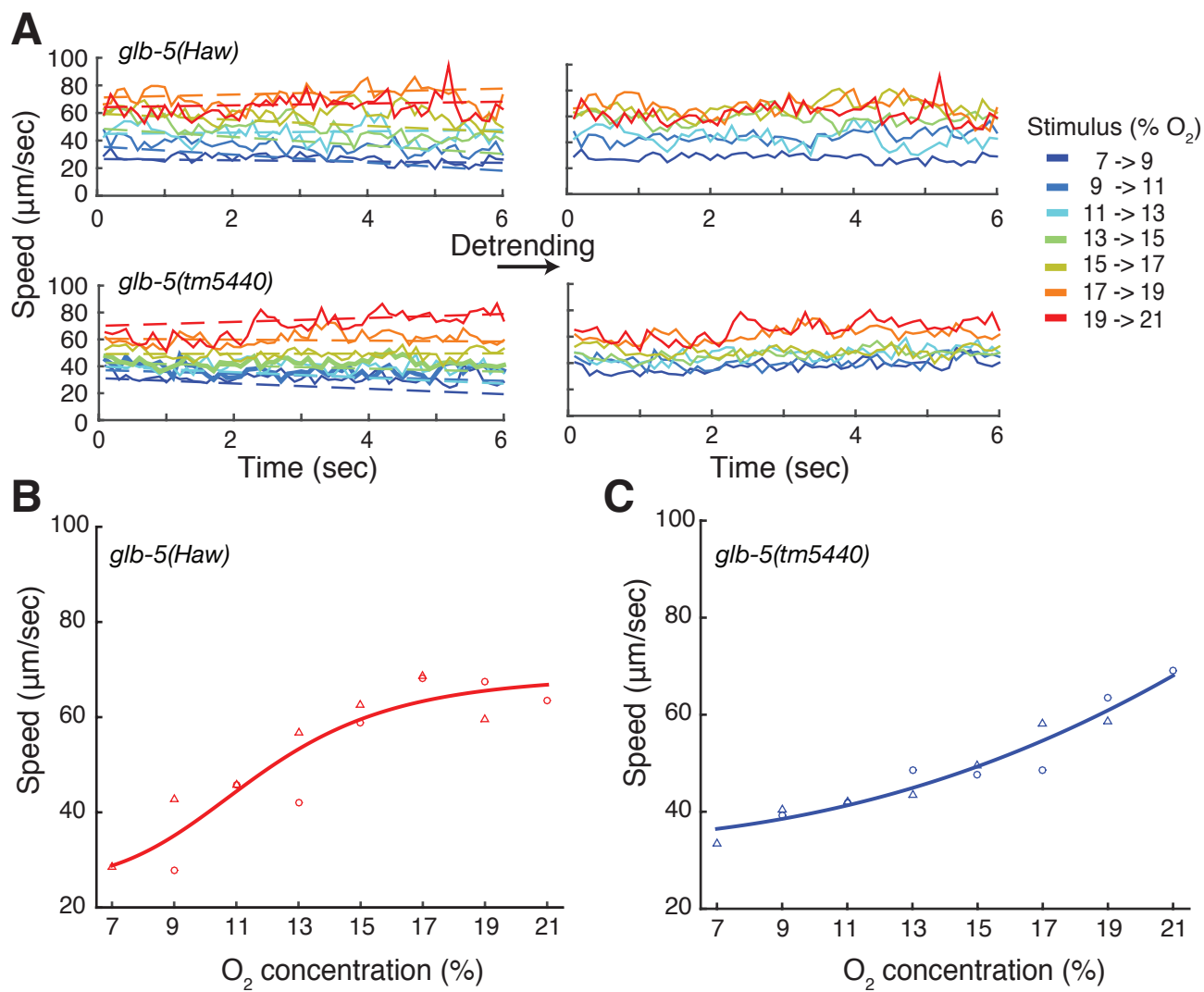

Fig S11

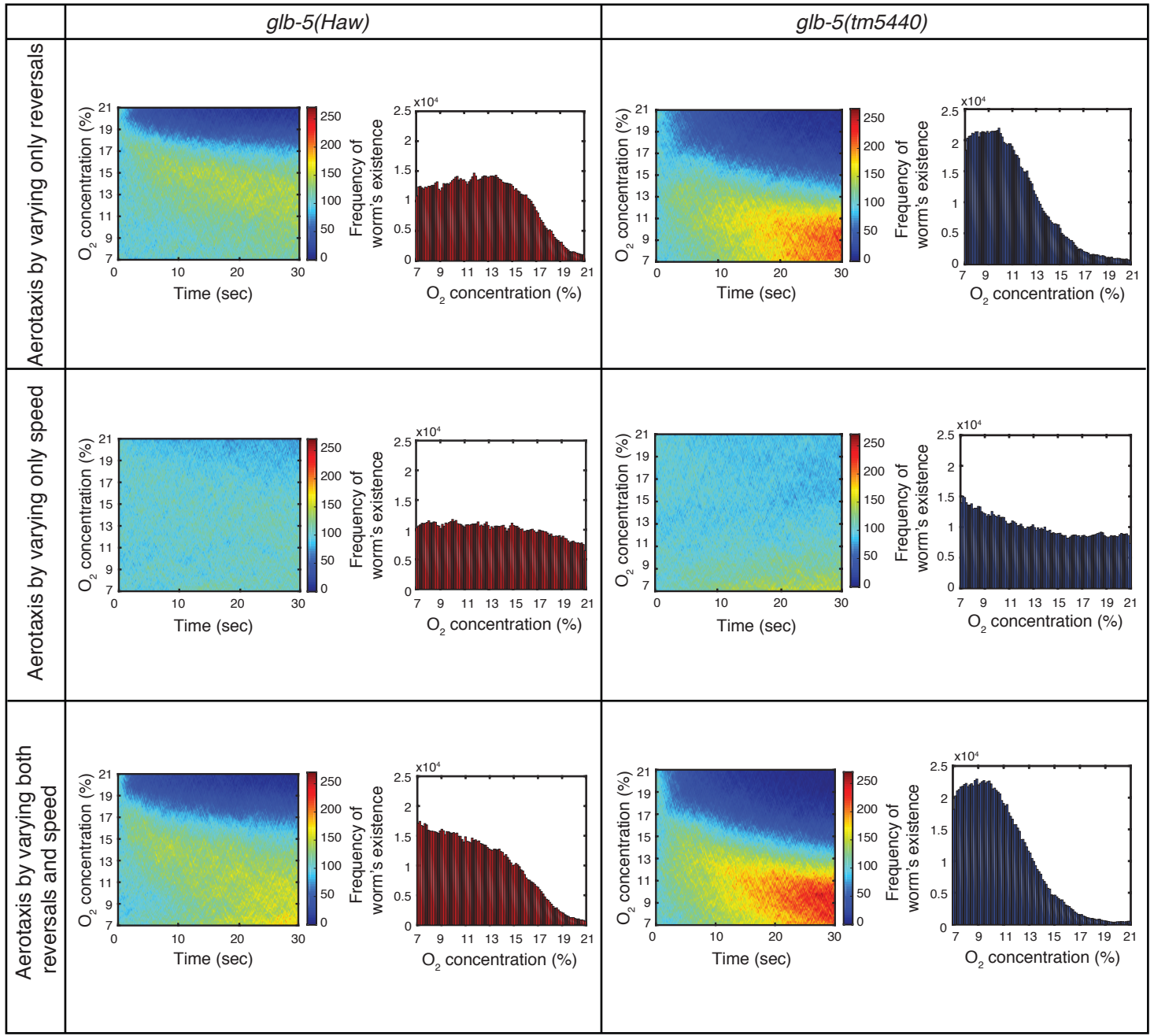